\documentclass[twocolumn,showpacs,preprintnumbers,amsmath,amssymb,noshowpacs]{revtex4}
\usepackage{graphicx}
\usepackage{dcolumn}
\usepackage{bm,color}

\begin{document}


\title{The production of $J/\psi$ associated with $c\bar{c}$ at ep colliders}



\author{Rong Li}
\address{School of Science, Xi'an Jiaotong University, Xi'an 710049,
China.}



\date{\today}

\begin{abstract}
We investigate the direct photoproduction of $J/\psi+c\bar{c}$ at $ep$ colliders with different center of mass energies in this paper. For the color-singlet part we take into account the contribution from QED subprocess as well as the QCD one. Our results show that the QED subprocess not only enhances the differential cross section in the large $p_t$ region but also gives a more transverse $J/\psi$ with $p_t$ increase, which is different from the almost unpolarized $J/\psi$ predicted by the QCD subprocess. The contribution from color octet subprocesses enhances the differential cross section by a factor of 5 in large $p_t$ region and predicts an even more transverse $J/\psi$ in the medium and large $p_t$ region. The sensitivity of this process to the color-octet matrix element $\langle O^{J/\psi}(^3S_1^8)\rangle$ may give great help in studying the matrix element.
\end{abstract}


\maketitle


\section{Introduction}\label{sec:introduction}
Quantum chromodynamics(QCD), the gauge theory determined by the non-abelian group, has been considered as the correct theory for describing the strong interaction among quarks and gluon. Because of the non-abelian property of the gauge group, QCD theory has asymptotic freedom and confinement. In view of the properties, we separate the related physical process into short distance parts which can be calculated order by order in the coupling constant and the non-perturbative long distance matrix elements (LDMEs) that are process independent and can be extracted experimentally or calculated by potential model or lattice QCD simulations. The heavy quarkonium, composed by charm(bottom) and anti-charm(bottom), provides a interesting physical system for us to study QCD that describes the interaction between the heavy quarks. For a long time the color singlet model (CSM) is used to describe the production and the decay of heavy quarkonium~\cite{Einhorn:1975ua}. In 1990s' there are huge conflicts between the experimental data~\cite{Abe:1992ww} and the theoretical predictions on the hadron-production of $J/\psi$ and $\psi'$. According to the non-relativistic characteristic of the heavy quarkonium a theory named non-relativistic quantum chromodynamics (NRQCD) was put forward~\cite{Bodwin:1994jh} which introduces the color-octet mechanism (COM) to fix the above conflicts and many works had been done to investigate the heavy quarkonium system within the NRQCD framework~\cite{Andronic:2015wma}.

The inclusive production of $J/\psi$ or $\Upsilon$ at hadron collider had been the hotspots for a long time. The theoretical calculations on their transverse momentum distribution and the polarization had been done at the next-to-leading order (NLO) for the color-singlet channel~\cite{Campbell:2007ws} and the color-octet channels~\cite{Ma:2010yw}. Even the partial next-next-to-leading order contribution had been estimated in reference~\cite{Artoisenet:2008fc}. It seems that combining the NLO QCD correction, the feed down contribution from higher quarkonium and the power correction could well describe the yield and polarization simultaneously on the hadroproduction of $J/\psi$ and $\Upsilon$. But the LDMEs extracted from inclusive hadronproduction of $J/\psi$ are not consistent with that extracted from $J/\psi$ production at B factories~\cite{Ma:2008gq}. Therefore, further studies are needed to clarify the mechanism on the production and decay of heavy quarkonium. Because the $v^2$ in $J/\psi$ meson is not very small ($v^2\approx0.3$) the relativistic correction should be considered carefully. The relativistic corrections on the color-singlet and the color-octet subprocess had been calculated in Refs.~\cite{Fan:2009zq} and gives the appreciable contribution to both yield and the polarization. Two years ago the soft gluon factorization scheme, a new framework that can handle this problem effectively, was raised~\cite{Ma:2017xno}. Ref.~\cite{Li:2019mdx} gives the relativistic correction to all orders for the related channels in the large $p_t$ limit. Those works may shed new lights on this problems. Moreover, we still want to exam the color octet mechanism and the property of the LDMEs in NRQCD from other aspects of heavy quarkonium physics. Therefore, the investigation has been extended to the associated production of heavy quarkonium, such as $J/\psi$ production associated with a heavy quark anti-quark pair. Many other aspects on the associated production can be find in the review paper~\cite{Lansberg:2019adr}. 

\begin{figure*}[ht]
\begin{center}
\includegraphics[scale=0.42]{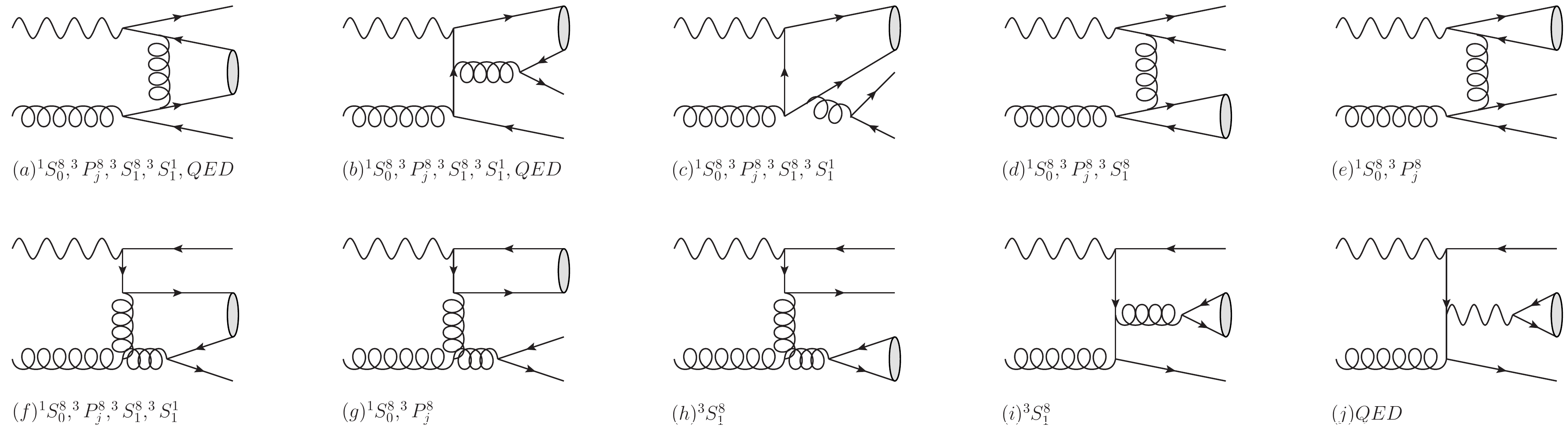}
\caption{\label{fig:feynmandiag} The typical feynman diagrams of the five subprocess for $J/\psi+c\bar{c}$ production.}
\end{center}
\end{figure*}

In 2002 the production of $J/\psi+c\bar{c}$ was investigated at B factory~\cite{Abe:2002rb} and lead to the severe conflict between experimental data and the leading order QCD calculations~\cite{Cho:1996cg}. About four years later the results of NLO QCD and relativistic corrections for this process reconcile the above conflict~\cite{Zhang:2006ay}. This combination of final states also had been investigated at $\gamma+\gamma$ collision~\cite{Qiao:2003ba}, or in the decay of Z boson~\cite{Barger:1989cq}. The hadroproduciton of $J/\psi+c\bar{c}$ was studied in reference~\cite{Baranov:2006dh,Artoisenet:2007xi}. In Ref.~\cite{He:2009zzb} the authors calculated the contribution from QED subprocess in the hadroproduction of $J/\psi+c\bar{c}$ within CSM and the numerical results indicate that the $p_t$ distribution is enhanced and the polarization also change drastically in the large $p_t$ region. The photoproduction of $J/\psi+c+\bar{c}$ in $ep$ collision is a similar process as that in hadron collider. In Ref.~\cite{Baranov:2006dh} the author had studied the yield of this process at HERA with both collinear parton model and the $k_t$ factorization approaches in color-singlet scheme. In 2013 the yield of $J/\psi$ at HERA had been considered again under the framework of NRQCD~\cite{Khan:2013ixa} and the numerical results indicate that the color-singlet contribution is comparable with the color-octet one at leading order in the lower $p_t$ region. Inspired by the studies in Ref.~\cite{He:2009zzb} we expect that the QED contribution may play an important role in the study of this process. In the following we investigate the yield and the polarization of $J/\psi$ in this process with NRQCD factorization formula and take both the QCD and the QED contributions into account for the color-singlet part in this process.

\section{Cross Section and Parameters}\label{sec:cross}

The cross section for the photoproduction of $J/\psi+c\bar{c}$ in NRQCD can be formulated as 
 \begin{eqnarray}\label{eq:cross}
&&\sigma(e+p \to J/\psi+c\bar{c}+X) = \int dx_1dx_2 \\
&\times& G_{\gamma/e}(x_1) G_{g/p}(x_2) \hat{\sigma}(\gamma+g \to
(c\bar{c})_n+c\bar{c}+X)\langle O^{J/\psi}_n\rangle. \nonumber
\label{eqn:factorization} \end{eqnarray}
Here  the $G_{g/p}(x_2)$ is the usual parton distribution function (PDF) of gluon and the $\langle O^{J/\psi}_n\rangle$ are the LDMEs describing the probability for an charm anti-charm configuration "n", which will be explained in detail later, evolving into $J/\psi$ in the long distance parts. The $G_{\gamma/e}(x_1)$ is the distribution function for a real $\gamma$ in electron which can be described by the Weizsacker-Williams (WW) approximation as follows~\cite{Williams:1934ad}:
\begin{equation}\label{WW} 
\begin{split}
G_{\gamma/e}(x)=&\frac{\alpha}{2 \pi}\left(2
m_e^2(\frac{1}{Q_{max}^2}-
\frac{1}{Q_{min}^2})x\right. \\ 
&\left. +\frac{(1+(1-x)^2)}{x}\log(\frac{Q_{max}^2}{Q_{min}^2})\right),
\end{split}
\end{equation}
where $x=E_{\gamma}/E_e$ is the energy fraction of the electron carried by the photon, $\alpha$ is the the fine structure
constant and $m_e$ is the electron mass. The definitions of
$Q_{max}^2$ and $Q_{min}^2$ are given by
\begin{eqnarray}
Q_{min}^2&=&\frac{m_e^2 x^2}{1-x},\\
Q_{max}^2&=&(\frac{\sqrt{s} \theta}{2})^2 (1-x)+Q_{min}^2,
\end{eqnarray}
where $\theta$ presents the angle between the momentums of photon and the electron which is taken as 32 mrad as in reference~\cite{Qiao:2003ba} which ensures the photon to be real. The $\hat{\sigma}$ in Eq.~(\ref{eq:cross}) is the cross section of the subprocess at the parton level. The direct photoproduction process only includes the photon gluon fusion subprocesses as following,
\begin{equation}\label{eq:subqcd}
\gamma+g \to c\bar{c}[^3S_1^1, ^1S_0^8, ^3P_j^8, ^3S_1^8]+c+\bar{c},
\end{equation}
where the spectroscopy notation $^{2s+1}L_j$ is used to indicate the $c\bar{c}$ configuration and the upper-right number 1 or 8 indicate the color state of the $c\bar{c}$ pair. In Eq.~(\ref{eq:subqcd}) the $^3S_1^1$ presents not only the QCD color-singlet subprocess mediated by a gluon but also the color-singlet subprocess mediated by a photon which is referred to as QED subprocess. We show the typical Feynman diagrams in Fig.~\ref{fig:feynmandiag} that were drawn by JaxoDraw~\cite{Binosi:2008ig}. Under every diagram we label the subprocesses that can be arose by the diagram. 

The $J/\psi$ polarization parameter $\alpha$ is defined as 
\begin{align}\label{eq:polar}
\alpha(p_t)=\frac{d\sigma_T/dp_t-2d\sigma_L/dp_t}{d\sigma_T/dp_t+2d\sigma_L/dp_t},
\end{align}
where $d\sigma_T$ and $d\sigma_L$ are the differential cross sections for transverse and longitudinal polarized $J/\psi$ respectively, $p_t$ is the transverse momentum of the $J/\psi$. 

\begin{figure*}[ht]
\begin{center}
\includegraphics[scale=0.30]{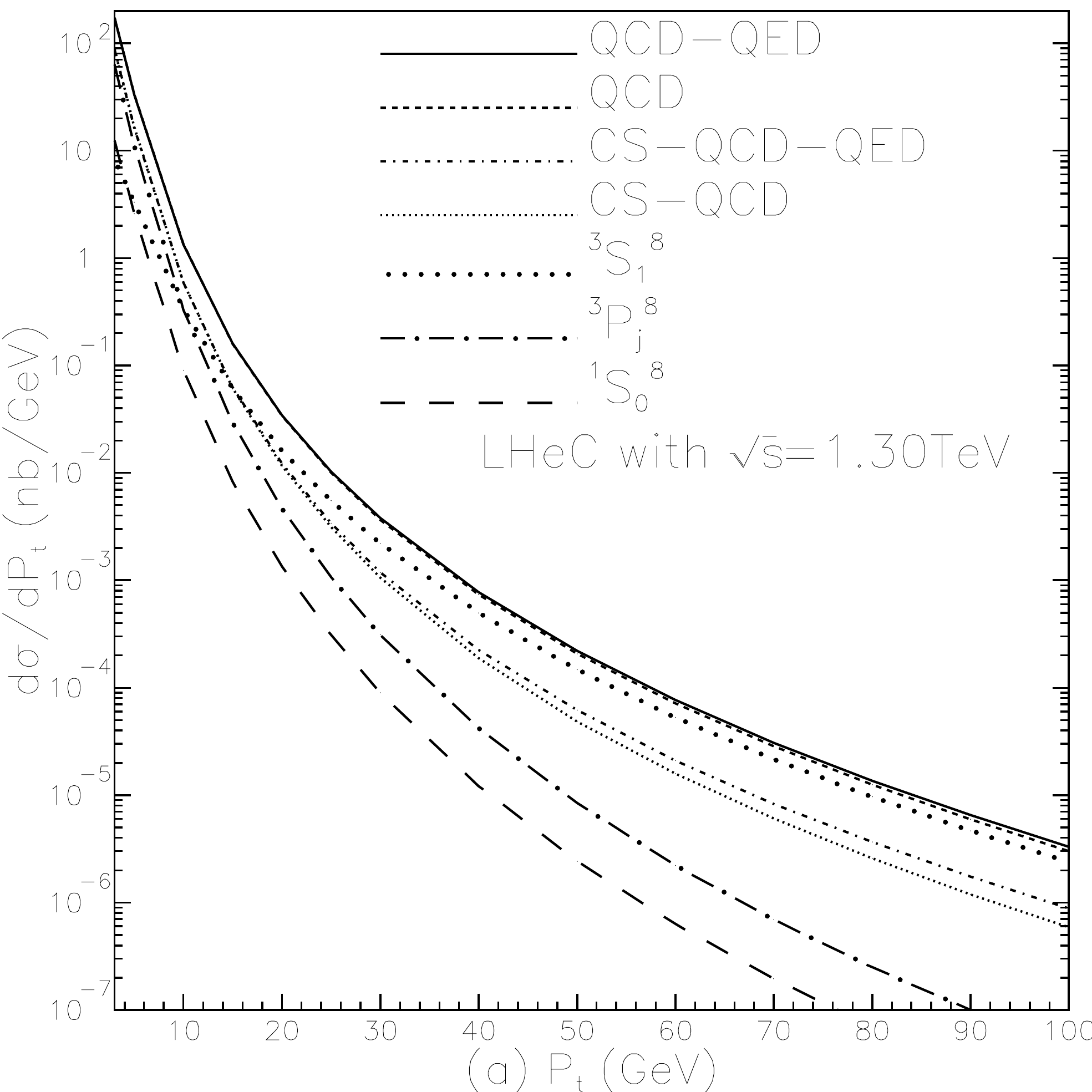}
\includegraphics[scale=0.30]{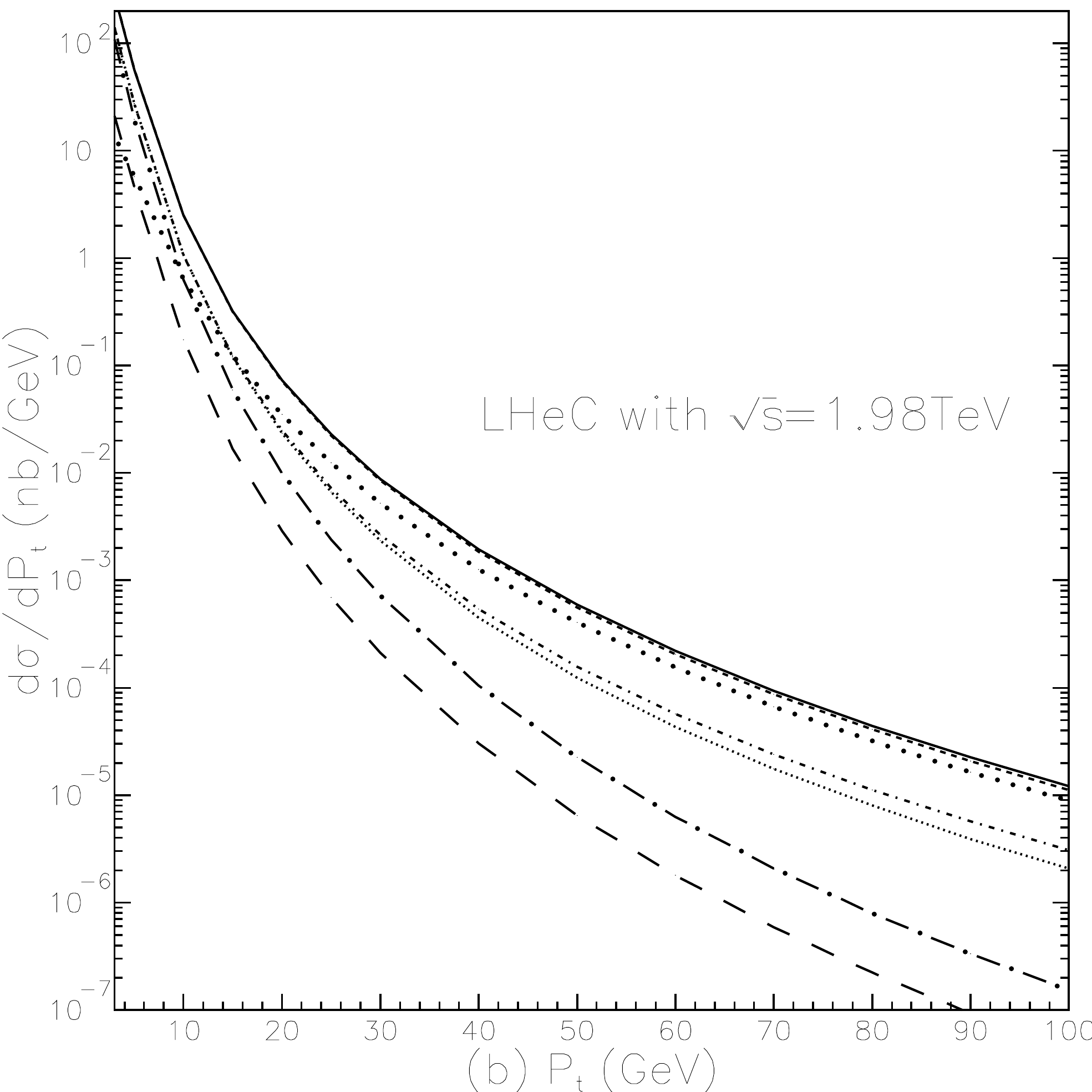}
\includegraphics[scale=0.30]{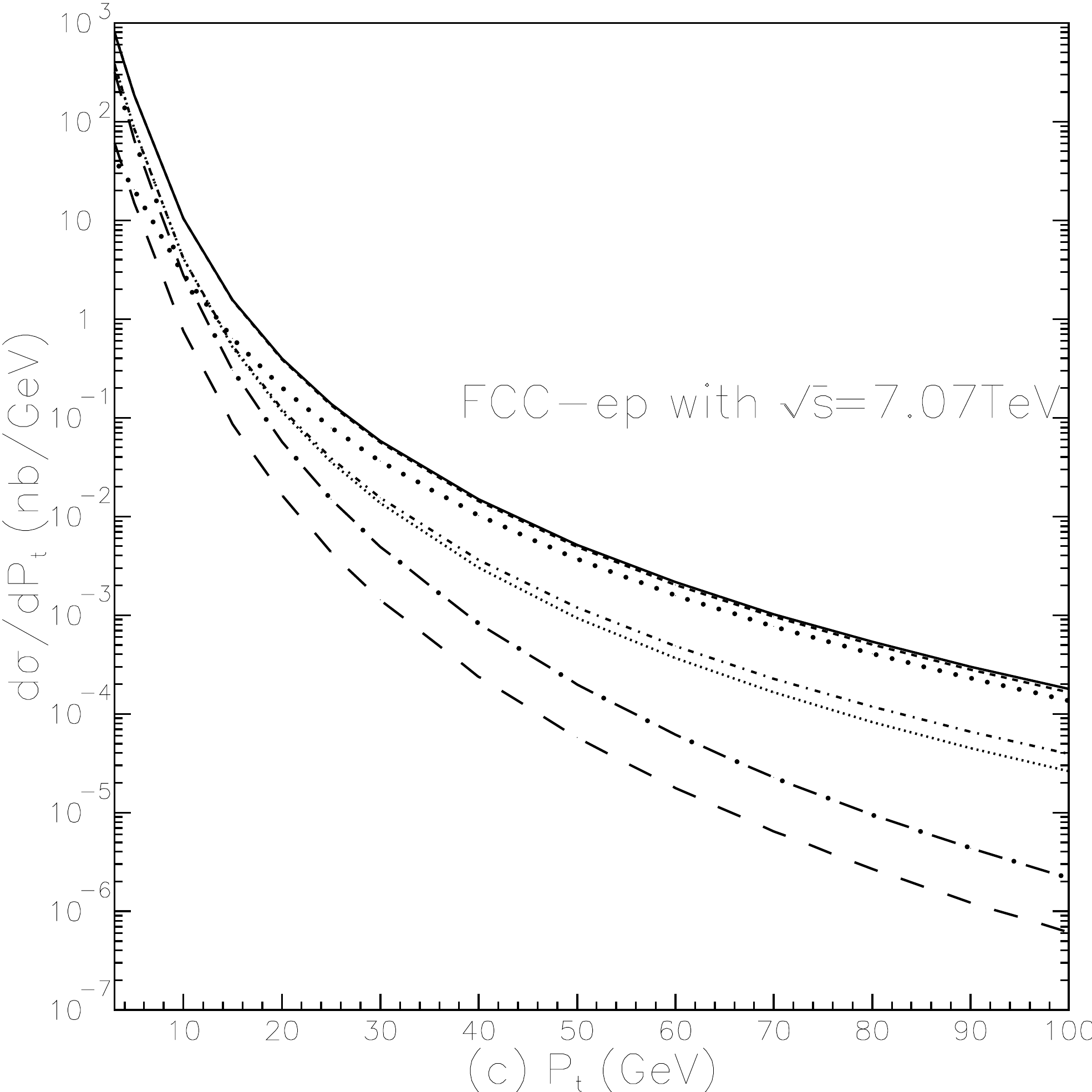}
\includegraphics[scale=0.30]{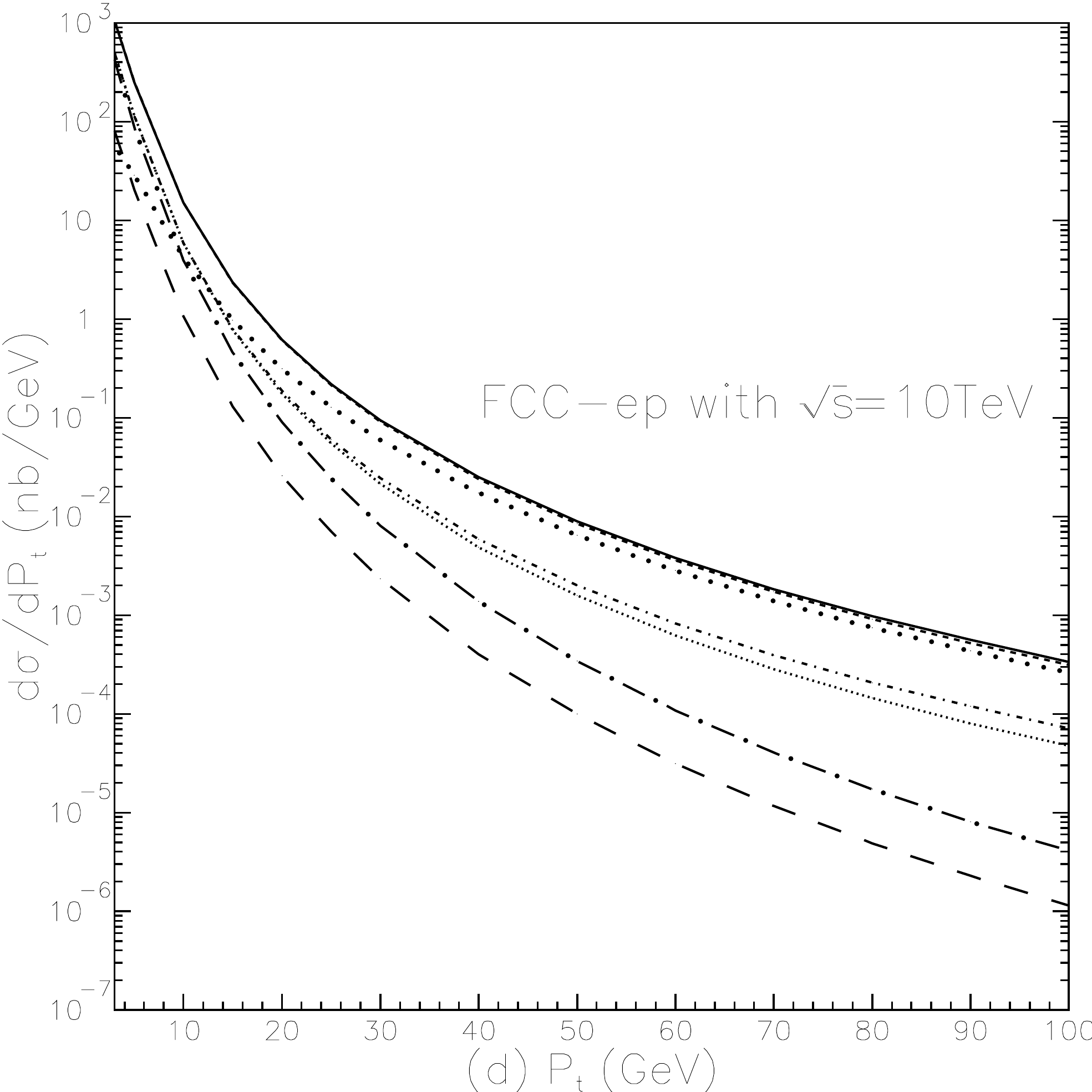}
\caption{\label{fig:ptdis} The $p_t(J/\psi)$ distributions for
$J/\psi+c\bar{c}$ production with different $\sqrt{s}$. The correspondence among the line types and the different subprocesses are indicated in the upper left figure (a). CS-QCD: result of CS QCD subprocess; CS-QCD-QED: result of CS QCD, QED subprocesses and the interference between them; QCD: result of CS and CO QCD subprocesses; QCD-QED: result of CS-QCD-QED and CO QCD subprocesses.}
\end{center}
\end{figure*}

We need calculate every subprocess and convolute them with the above two distribution functions. We use the Feynman Diagram Calculation (FDC) package~\cite{Wang:2004du} to generate the Feynman amplitude and the phase space integration in the fortran form. By choosing the proper values for the relevant parameters we can get the numerical results. 

The charm quark mass is set as $m_c=$1.5GeV. The factorization and the renormalization scale are chosen as $\mu_f=\mu_r=\sqrt{4m_c^2+p_t^2}$. The fine structure constant is set as $1/128$. The CT14LL parton distribution function and the corresponding one-loop running coupling constant~\cite{Dulat:2015mca} are used to convolute with the parton level results. As for the LDMEs, we use the leading order results as following
\begin{align}\label{eq:LDME}
\langle O^{J/\psi}(^3S_1^1)\rangle&=1.1 \text{GeV}^3,\\
\langle O^{J/\psi}(^1S_0^8)\rangle&=1\times10^{-2}\text{GeV}^3,\\
\langle O^{J/\psi}(^3S_1^8)\rangle&=1.12\times10^{-2} \text{GeV}^3,\\
\langle O^{J/\psi}(^3P_0^8)\rangle&=11.25\times10^{-3}\text{GeV}^5,
\end{align}
where the color-singlet one is extracted from $J/\psi \to \mu^+\mu^-$and the color-octet ones are extracted from fitting the CDF data~\cite{Fleming:1997fq}. In the future, there are two planed large electron-proton collider projects. One is the the Large Hadron Electron Collider (LHeC) project~\cite{AbelleiraFernandez:2012cc} and the other is the Multi-TeV center of mass energy ep colliders based on the Future Circular Collider (FCC-ep)~\cite{Acar:2016rde}. The corresponding beam energies and the center of mass energies are listed in Tab.~\ref{tab:energy}. 

\begin{table}[h]
\caption[]{The beam energies and the center of mass energies of the $ep$ colliders in unit of TeV.)}
\label{tab:energy}
\renewcommand{\arraystretch}{1.5}
\[
\begin{array}{|c|c|c|c|}
\hline \hline  &E_e&E_p&\sqrt{s} \\
\hline \text{LHeC1}&0.060&7&1.30\\
\hline \text{LHeC2}&0.140&7&1.98\\
\hline \text{FCC-ep1}&0.250&50&7.07\\
\hline \text{FCC-ep2}&0.500&50&10.00\\
\hline \hline
\end{array}
\]
\renewcommand{\arraystretch}{1.0}
\end{table}

\section{Numerical results}\label{sec:num.}

\begin{figure*}
\begin{center}
\includegraphics[scale=0.30]{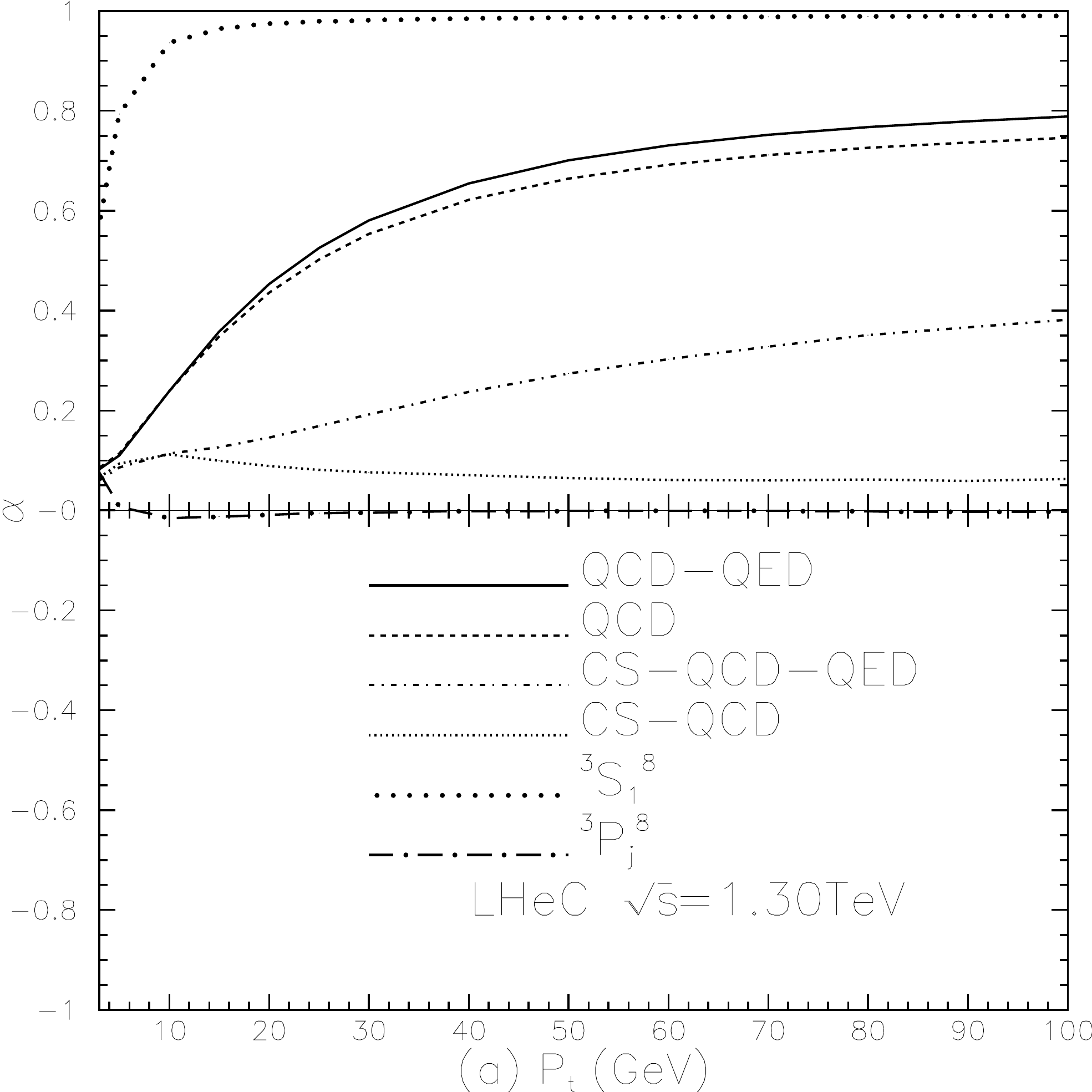}
\includegraphics[scale=0.30]{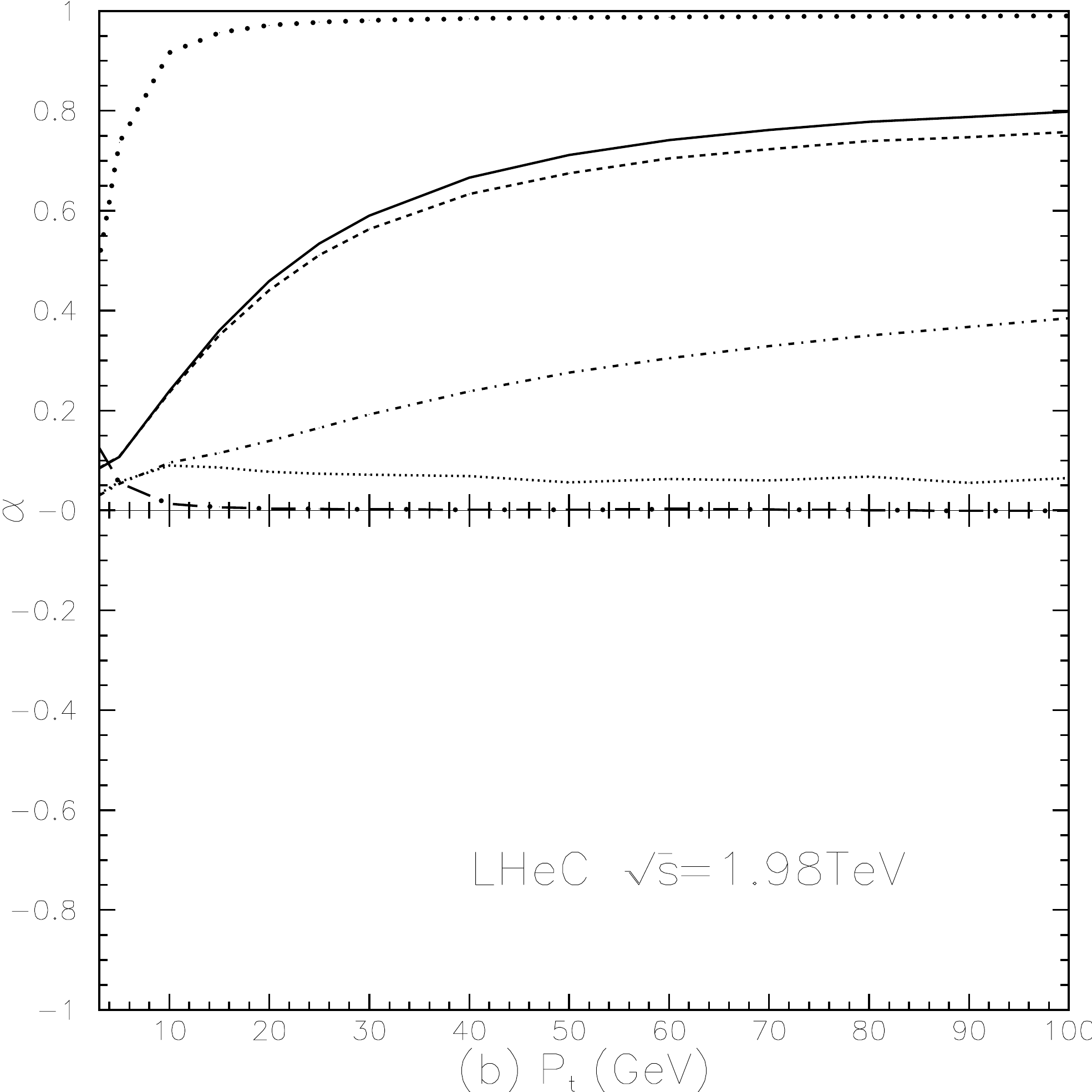}
\includegraphics[scale=0.30]{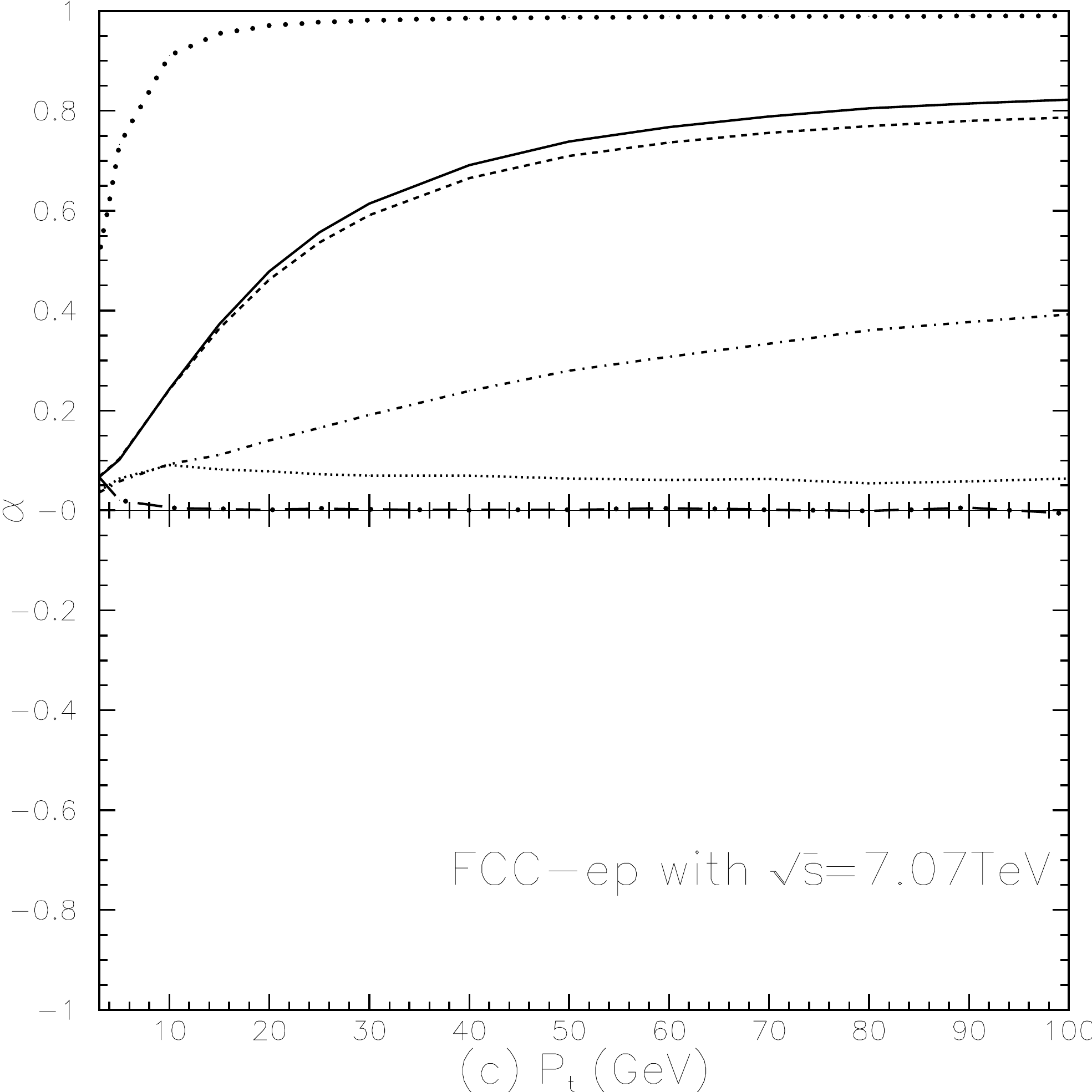}
\includegraphics[scale=0.30]{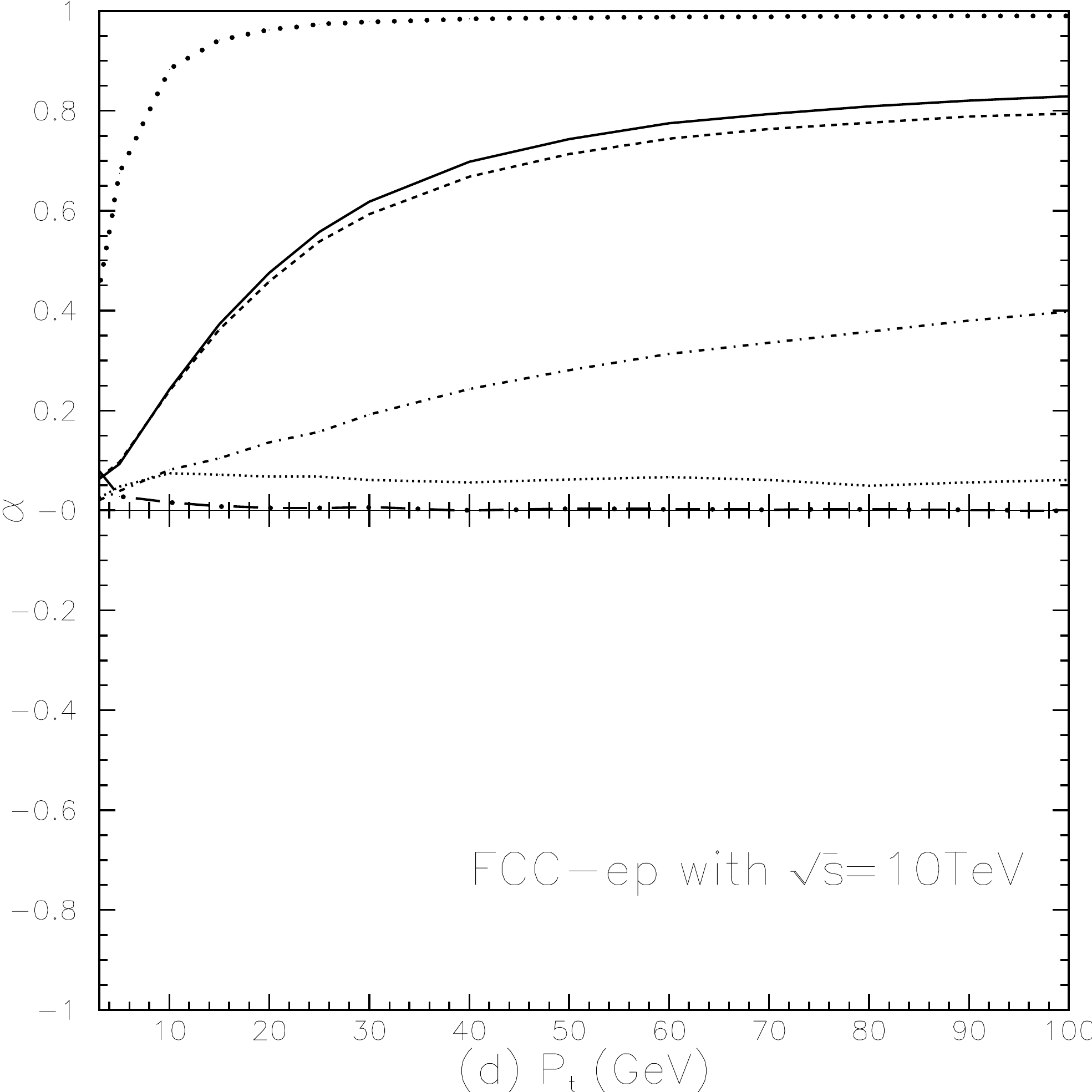}
\caption{\label{fig:pldis} The polarization distributions for
$J/\psi$ production associated with a $c\bar{c}$ pair. The correspondence among the line types and the different subprocesses are indicated in the upper left figure (a).}
\end{center}
\end{figure*}

We present the $p_t$ distribution of $J/\psi$ in the Fig.~\ref{fig:ptdis}. It can be seen that the differential cross section for the QED subprocess is very tiny in the small $p_t$ region and appreciable in the large $p_t$ region. The reason is similar as that in reference~\cite{He:2009zzb}. The contribution of QED subprocess is suppressed by $(\alpha/\alpha_s)^2$ comparing to the QCD one in small $p_t$ region. As to the large $p_t$ region the kinematic enhancement from the photon fragmentation process makes the yield of QED subprocess non-negligible. Therefore, the QED subprocess should be ignorable in the total cross section calculation and be taken into account in the $J/\psi$ distribution in large $p_t$ region for color-singlet contribution. The decrease of $\alpha_s$ with the increase of $p_t$ even promotes the contribution of the QED subprocess. The numerical results also show that the color octet subprocess $\gamma+g \to c\bar{c}[^3S_1^8]+c+\bar{c}$ gives the dominate contribution in the large $p_t$ region. The total results are about 5$\sim$6 times larger than the QCD CS ones with different $\sqrt{s}$. Accordingly, the $p_t$ distribution in the large $p_t$ region should be sensitive to the value of the matrix element $\langle O^{J/\psi}(^3S_1^8)\rangle$ and will provide us a new opportunity to study this color octet matrix element. In the lower $p_t$ region the main contribution comes form the $^3S_1^1$ and the $^3P_j^8$ subprocesses. Considering the difference among the LDMEs our results on the $J/\psi$ distribution in the lower $p_t$ region are consistent with that of Ref.~\cite{Khan:2013ixa}.

In Fig.~\ref{fig:pldis} the distributions of polarization parameter $\alpha$ as the function of transverse momentum of $J/\psi$ are plotted. As for the color-singlet channels, the QCD subprocess predicts the small transverse polarization in the whole $p_t$ region. By including the contribution from the QED subprocess the $J/\psi$ polarization changes to more and more transverse with $p_t$ increase and tends to 0.4 at $p_t=100$GeV, which is because the photon fragmentation subprocess yields the transverse polarized $J/\psi$ and gives the appreciable contribution in the large $p_t$ region. That is to say, the nearly unpolarized color-singlet QCD subprocess yields a small numerator in calculating polarization parameter $\alpha(p_t)$ and the QED subprocess makes the numerator increasing with the increase of $p_t$. We do not plot the polarization of $^1S_0^8$ subprocess because it contributes the unpolarized $J/\psi$. The $^3P_j^8$ subprocess is almost unpolarized in the whole $p_t$ region and the $^3S_1^8$ subprocess is almost transverse in the medium and large $p_t$ region. Because the contributions from $^1S_0^8$ and $^3P_j^8$ subprocesses decrease rapidly with the increasing of $p_t$, which comes from their $1/p_t^6$ behavior, they do not affect the $J/\psi$ polarization in the large $p_t$ region practically. Therefore, when we take the color-octet contribution into account, especially the $^3S_1^8$ subprocess, the polarization of $J/\psi$ takes a considerably change to more transverse in the medium and large $p_t$ region. It can be seen that the differential cross section for both $^3S_1^1$ and $^3S_1^8$ subprocesses all behave as $1/p_t^4$ and the gap between them is sensitive to the value of matrix element $\langle O^{J/\psi}(^3S_1^8)\rangle$. The differential cross section shows that only at the large $p_t$ region the $^3S_1^8$ subprocess will give the dominant contribution and can be manifestly distinguished from the color-singlet one. Therefore, combining the $p_t$ distribution and the polarization studies at $ep$ colliders may give a well constrain on the color-octet matrix element $\langle O^{J/\psi}(^3S_1^8)\rangle$.

The design luminosity of LHeC and the FCC-ep are $10^{33}\text{cm}^{-2}\text{s}^{-1}$~\cite{AbelleiraFernandez:2012cc} and $10^{32}\text{cm}^{-2}\text{s}^{-1}$~\cite{Acar:2016rde} respectively. According to our numerical results there are about $10^2$ ($10^3$) events at $p_t=$100GeV and $6\times10^3$ ($2.7\times10^4$) events at $p_t=$50GeV at LHeC (FCC-ep). Considering that the charm quark tagging efficiency is about $41\%$~\cite{Aaboud:2018fhh} and the brach ration of $J/\psi \to e^+e^- (\mu^+\mu^-)$ is about $12\%$~\cite{Tanabashi:2018oca} we expect there are still enough events in the medium and large $p_t$ regions. Combining the di-lepton decay of $J/\psi$ and the tagging of charm quark may suppress the backgrounds and makes this process more accessible.


\section{Summary}\label{sec:summary}

In summary we investigate the photoproduction of $J/\psi+c\bar{c}$ in future $ep$ colliders with different center of mass energies. In our analysis we not only consider the color-singlet and color-octet channels in QCD but also take the QED subprocess into account. Both the physical analysis and the numerical results indicate that the color-singlet QED subprocess, especially the photon fragmentation mechanism, is important for the studies on $J/\psi$ distribution and polarization within the CSM. It enhances the differential cross section of color-singlet QCD subprocess by a factor of 0.5 and changes the $J/\psi$ polarization from almost unpolarized to more transverse with the increasing of $J/\psi$ transverse momentum. Our calculations demonstrate that the color-octet subprocesses give the large contribution to photoproduciton of $J/\psi+c\bar{c}$ in large $p_t$ region and the $^3S_1^8$ channel is the dominate one. The total differential cross section is about 5 times larger than that of color-singlet QCD subprocess and the polarization of $J/\psi$ becomes more transverse in the medium and large $p_t$ region. The numerical results of this process are sensitive to the value of matrix element $\langle O^{J/\psi}(^3S_1^8)\rangle$. Therefore, to investigate it in the future $ep$ colliders will help us to study the color octet matrix element $\langle O^{J/\psi}(^3S_1^8)\rangle$. Owing to the heavier mass and the small electric charge, this photon fragmentation mechanism would give less contribution in $\Upsilon$ case.

\begin{acknowledgments}

We acknowledge the supports from the National Natural Science Foundation of China under Grants No. 11375137 and U1832160, the Natural Science Foundation of Shaanxi Province under Grants No. 2015JQ1003 and the Fundamental Research Funds for the Central Universities.

\end{acknowledgments}


\end{document}